\documentclass[10pt,aps,prb,raggedbottom,longbibliography,reprint,citeautoscript,letterpaper,bibnotes]{revtex4-2} 
\usepackage[usenames,dvipsnames]{color}
\usepackage{graphicx,microtype}
\usepackage[bookmarks=false,colorlinks]{hyperref}
\hypersetup{linkcolor=magenta,citecolor=MidnightBlue,filecolor=Plum,urlcolor=MidnightBlue}
\usepackage[all]{hypcap} 
\usepackage{lmodern}
\usepackage{amsmath,mathrsfs,cleveref}
\usepackage{soul}

\newcommand{\pgp}[1]{\textcolor{black}{#1}}



\begin{document}

\preprint{APS/123-QED}

\title{Transition-state lattice modes and the breakdown of adiabatic tunneling for hydrogen and deuterium in bcc Nb}

\author{P. Graham Pritchard}
\author{James M.\ Rondinelli}%
\email{jrondinelli@northwestern.edu}
\affiliation{
Department of Materials Science and Engineering, Northwestern University, Evanston, Illinois 60208, USA
}

\date{\today} 

\begin{abstract}
Interstitial hydrogen and deuterium in body-centered-cubic metals constitute archetypal quantum tunneling systems.
Their relevance has been renewed by the connection between hydrogenic tunneling in Nb and defect-induced decoherence in superconducting qubits, motivating a predictive microscopic theory.
Existing theoretical treatments invoke an adiabatic separation between the light interstitial and the host lattice, an assumption whose validity has not been rigorously established for hydrogenic species.
Here, we show that the experimentally measured tunnel splittings of
O-trapped H and D in bcc Nb are quantitatively well-described within a five-dimensional (5D) Lattice-Renormalized Born-Oppenheimer (LRBO) framework. This approach treats three interstitial modes and two judiciously selected lattice modes, which includes a transition-state mode, on equal quantum footing.
By recasting nested Born-Oppenheimer hierarchies within this same formalism and
benchmarking against modern \textit{ab initio} potential energy surfaces, we show
that adiabatic separation of the light particle from lattice dynamics is satisfied
only in the positive-muon ($\mu^{+}$) mass limit. In contrast, tunneling for H and D is
fundamentally a collective, nonadiabatic process mediated by anharmonic
lattice couplings.
Finally, we show that the breakdown of adiabaticity can be anticipated from
simple energy estimates involving the ground-state light-particle energy evaluated
at a small number of fixed lattice configurations, providing a practical criterion
for assessing the validity of adiabatic tunneling theories in other systems.
\end{abstract}

\maketitle

\section{\label{sec:level1}Introduction}
Light interstitials such as hydrogen and deuterium form quantum tunneling systems in crystalline solids, giving rise to low-temperature anomalies in thermodynamic and dynamical responses. 
In body-centered-cubic Nb, interstitial H and D exhibit tunneling behavior $\le2$\,K when trapped by secondary interstitial impurities (C, N, O) or substitutional solutes (Ti, Zr). 
This behavior has been established from the anomalous, non-Debye contributions to the specific heat \cite{Sellers1973, Neumaier1982, Magerl1983, Wipf1984p2}, an energy loss peak in inelastic neutron scattering \cite{Wipf1984}, and anelastic-relaxation peaks \cite{Mattas1975, Cannelli1994}. 
In the case of O-trapped H and D \cite{Wipf1984, Wipf1984p2}, these observations fit well to effective two-level system models, yielding tunnel splittings $J_\mathrm{H}=0.19$ meV for H and one order of magnitude smaller $J_\mathrm{D}=0.021$ meV for D.

Hydrogenic tunneling systems in Nb are not only of fundamental interest but are increasingly relevant to superconducting technologies.
Baths of such tunneling systems \pgp{have been theoretically predicted} to generate subgap quasiparticle states in BCS superconductors, such as Nb \cite{He2025}, contributing to excess dissipation and enhanced decoherence rates in superconducting qubits \cite{Catelani2012, Glazman2021, Siddiqi2021}.
Developing an accurate and predictive microscopic description of tunneling systems is therefore essential for identifying materials-based mitigation strategies for defect-induced quasiparticles.

Despite this importance, previous models of hydrogen tunneling in bcc Nb have been framed in an adiabatic, light-interstitial picture in which the light particle tunnels through a static \cite{Sundell2004, Abogoda2025b} or quasi-static lattice potential \cite{Sugimoto1980}. 
Within the framework proposed by Sugimoto et al.~\cite{Sugimoto1980}, the wavefunction of H or D is assumed to respond instantaneously relative to the motion of the lattice allowing lattice forces to be computed using the Hellman-Feynman theorem.
Although this approximation has proven useful in some contexts, and may be formally justified for extremely light particles such as positive muons ($\mu^+$), its validity for hydrogenic species in real, anharmonic host lattices remains largely unverified.
Recently, we have begun to expose the limitations of this approach. 
Using a lattice-renormalized tunneling formalism that treats the light interstitial and select lattice mode(s) on equal quantum footing \cite{Pritchard2025b}, we previously showed that the tunnel splittings of O-trapped H and D in Nb are bounded by carefully selecting low-dimensional,  3D and 4D subspaces, of the full nuclear Hamiltonian. 
These results already indicated that lattice participation plays a non-perturbative role in determining tunneling energetics. 
Two key questions, however, remain unresolved: what level of lattice coupling is required for quantitative agreement with experiment, and whether effective two-level descriptions of tunneling systems (TS) remain valid for different chemical species?
In this Article, we answer both questions and demonstrate a breakdown of the adiabatic tunneling paradigm for hydrogenic interstitials in Nb. 
We show that the experimentally observed tunnel splittings of O-trapped H and D are quantitatively \pgp {well-described} when the tunneling problem is formulated in a five‑dimensional (5D) subspace that explicitly incorporates coupled lattice modes, establishing a controlled and converged description of lattice-renormalized tunneling.
Next, because the same interstitial H sites may be occupied by positive muons, $\mu^+$,  whose mass is approximately 9-fold smaller than that of a proton, we critically assess the validity of the adiabatic light-interstitial hypothesis by systematically examining the mass dependence of tunneling from $\mu^+$ to H and D. 
We show that only $\mu^+$ satisfies the conditions required for adiabatic separation from the lattice \cite{Sugimoto1980}, whereas the tunneling dynamics of H and heavier isotopes are fundamentally collective, mediated by anharmonic couplings to lattice degrees of freedom. 
We additionally show the applicability of this formalism to compute ``coincidence'' structures in a self-consistent manner and show that it is equivalent to the minimization of the lattice with an equal population constraint of the degenerate minima. 
Together, these results establish hydrogen tunneling in superconducting Nb as a lattice-mediated, multilevel quantum phenomenon with direct implications for defect-induced decoherence in superconducting materials.

\section{Results and Discussion}

\subsection{Lattice-renormalized model}
\begin{figure}[t]
\centering\vspace{-6pt}
   \includegraphics[width=0.45\textwidth]{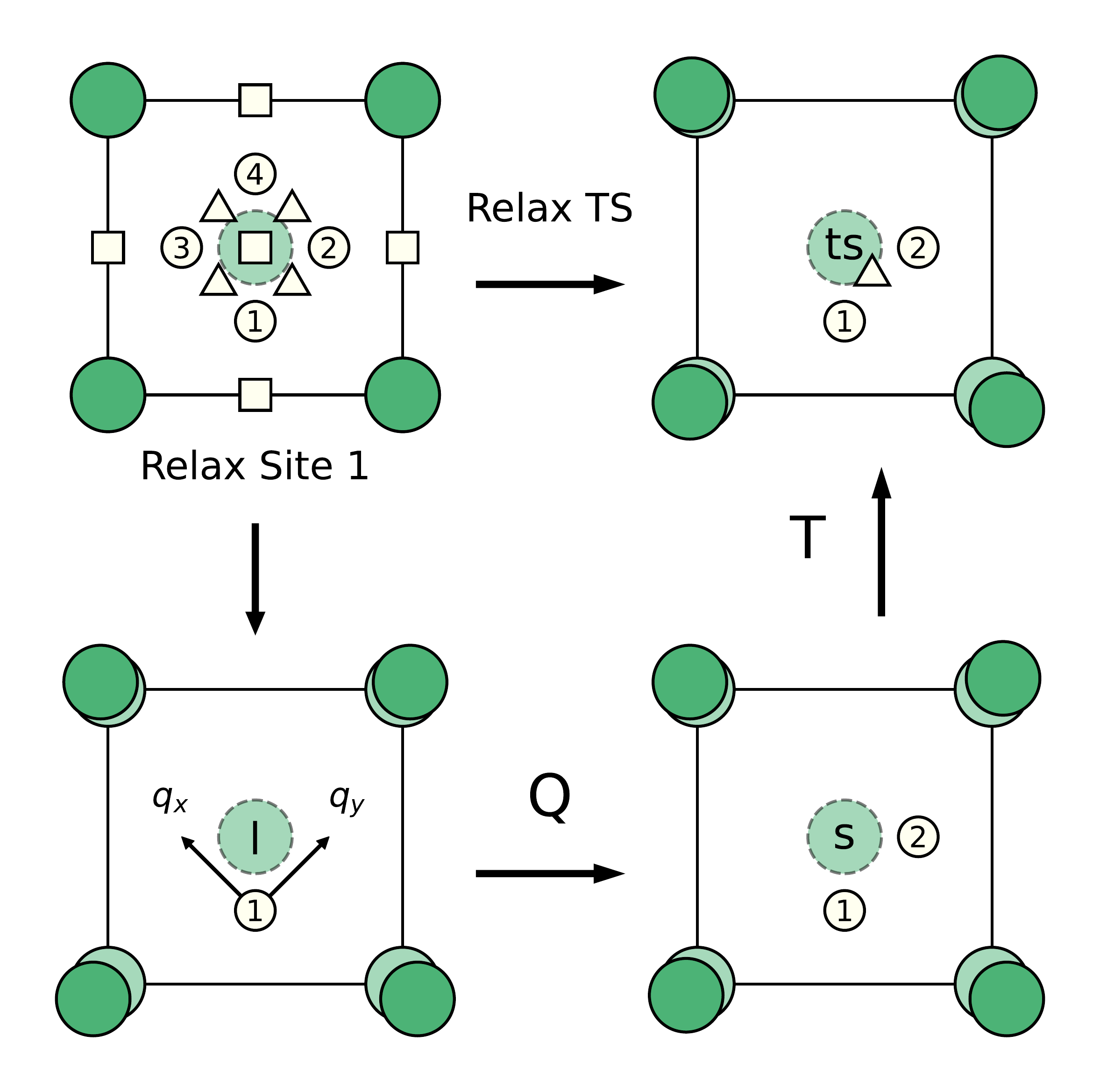}
   \vspace{-12pt}
   \caption{
   Schematic illustration of lattice order parameters defining the tunneling subspace.
   Nb atoms (green), degenerate tetrahedral H sites (white circles), and degenerate saddle points    (white triangles) and octahedral sites (white squares) are shown within fixed lattice configurations; displaced Nb positions indicate lattice relaxation.
   Atoms with solid outlines lie in the same plane, while the dashed Nb atom is body-centered.
   The order parameter $Q$ transforms a singly degenerate hydrogen configuration ($l$) into a twofold-degenerate symmetric configuration ($s$) and
subsequently into a second singly degenerate configuration ($r$) that is not shown.
   The order parameter $T$ distorts the symmetric configuration toward the transition state configuration ($ts$) located at the barrier maximum along the minimum-energy path.}
   \label{fig:h_sites}
\end{figure}

The tunneling behavior of defect-trapped H in Nb originates from a small number of degenerate hydrogen configurations linked together by small distortions of the Nb lattice.
As shown in \pgp{Ref.}~\cite{Pritchard2025b}, both interstitial and substitutional defects generate distinct manifolds of symmetry-related tetrahedral sites that form the microscopic basis for low-temperature tunneling systems.
A quantitative description of tunneling in these systems, therefore, requires a Hamiltonian that explicitly incorporates the coupled motion of hydrogen and the local lattice degrees of freedom that mediate transitions between degenerate configurations.
To that end, we introduced a lattice-renormalized formalism to compute tunnel splittings for such configurational tunneling systems \cite{Pritchard2025b}. 
The minimal subspace of the nuclear Hamiltonian, $\hat H_n^{sub}$, includes three hydrogen modes, $\mathbf{q}$, describing the motion between adjacent tetrahedral sites, together with a single lattice mode, $Q$, that transforms the lattice between configurations with degenerate hydrogen tetrahedral sites: 
\begin{equation}\label{eq:reduced}
    \hat H_n^{\text{sub}} = \sum_{i=\mathbf{q},Q}\frac{-\hbar^2}{2}\nabla_i'^2 + V(\mathbf{q},Q)\,,
\end{equation}
where all real-space coordinates have been transformed to mass-normalized phonon coordinates.
This construction captures the leading lattice degree of freedom associated with tunneling but implicitly constrains the lattice to follow a direct path between degenerate sites. 
We refer to this Hamiltonian and the extension described below as Lattice-Renormalized Born-Oppenheimer (LRBO) approximations. 
Here, we relax this constraint by introducing an additional lattice mode, $T$, which parametrizes distortions toward the transition-state configuration: the lattice arrangement corresponding to the barrier maximum along the minimum-energy tunneling pathway.
The augmented Hamiltonian then takes the form:
\begin{equation}\label{eq:reduced_5d}
    \hat H_n^{\text{sub}} = \sum_{i=\mathbf{q},Q, T}\frac{-\hbar^2}{2}\nabla_i'^2 + V(\mathbf{q},Q, T)\,.
\end{equation}
The lattice coordinates $Q$ and $T$ are constructed from three fully relaxed atomic configurations.
Two structures correspond to hydrogen occupying adjacent, degenerate-by-symmetry tetrahedral sites (labeled Site 1 and 2 in \autoref{fig:h_sites}), while the third structure corresponds to the unstable equilibrium configuration located midway between these sites (marked by a \pgp{triangle}), which possesses a single symmetry-protected unstable mode.
This structure, therefore, defines the transition state between Sites 1 and 2.
Denoting the relaxed atomic coordinates (defined relative to the center-of-mass of the lattice without H) as $\mathbf{R}^{l}$, $\mathbf{R}^{r}$, and $\mathbf{R}^{ts}$, we define
\begin{align}
\mathbf{Q}&=M^{1/2}\left(\mathbf{R}^{r}-\mathbf{R}^{l}\right),\\  
\mathbf{T}&=M^{1/2}\left(\mathbf{R}^{ts}-\tfrac{1}{2}\mathbf{R}^{l}-\tfrac{1}{2}\mathbf{R}^{r}\right)\,,
\end{align}
where $M$ is the mass matrix defined in Ref.~\cite{Pritchard2025b}.
We then solve \autoref{eq:reduced_5d} numerically, as described in Appendix \ref{sec:calcEq2}, with inputs from density functional theory (DFT) simulations.

\begin{figure*}
\centering
   \includegraphics[width=0.49\textwidth]{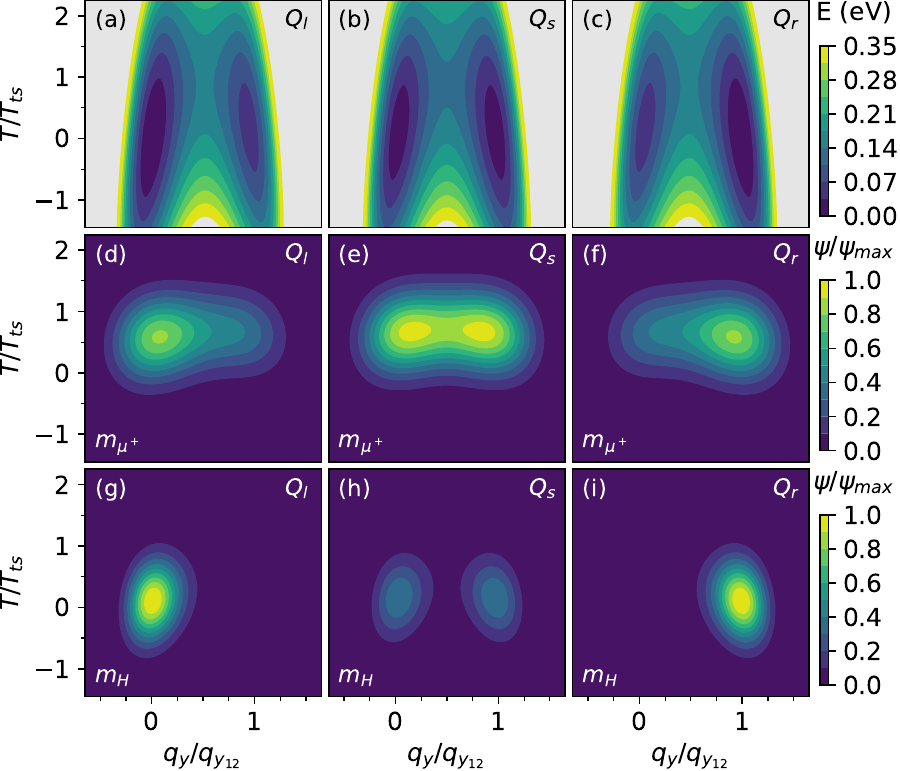}
    \includegraphics[width=0.49\textwidth]{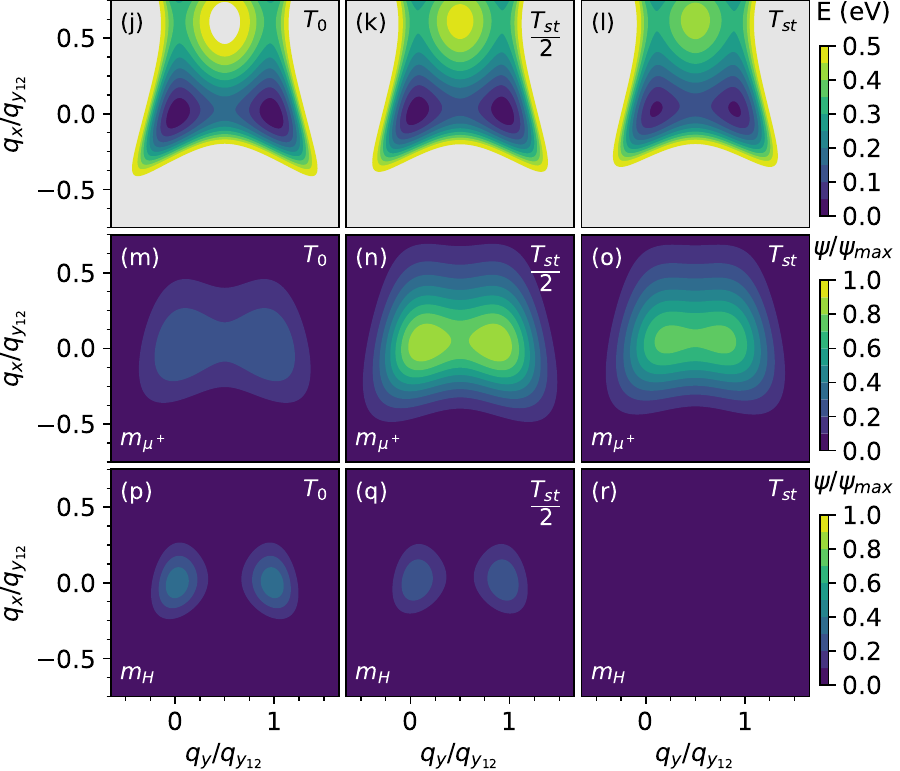}
   \caption{Contour plots of the potential energy surface and ground-state wavefunction for two light particle masses: $m=m_{\mu^+}$ and $m=m_\mathrm{H}$. (a-i) Slices  along the $q_y$ and $T$ modes with $q_x=q_z=0$, evaluated at three representative lattice configurations: (a, d, g) $Q=Q_l$, (b, e , h) $Q=Q_s$, (c, f, i) $Q=Q_r$. (j-r) Slices along the $q_x$ and $q_y$ modes with $q_z=0$ and $Q=Q_s$, shown at three values of the transition-state coordinate: (j, m p) $T=T_0=0$, (k, n, q) $T=T_{ts}/2$, and (l, o, r) $T=T_{ts}$. Wavefunctions are normalized by their most probable value. Cubic spline interpolation is used to smoothly render the potential energy and wavefunction contours.}
   \label{fig:oh_vphi}
\end{figure*}

Solving \autoref{eq:reduced_5d}, we obtain tunnel splittings $J=E_1-E_0$ of 0.13 meV (0.012 meV) for O-trapped H (D), in good agreement with the experimentally reported values of 0.19 meV (0.021 meV). These results were computed at a H concentration of $1/54$. Structural details of the stable H sites are described in \pgp{Ref.}~\cite{Pritchard2025b}.
These values are substantially improved over the lower bounds of 0.064 meV ($4.8\times10^{-3}$ meV) reported in \pgp{Ref.}~\cite{Pritchard2025b}, demonstrating the quantitative importance of explicitly resolving the transition-state lattice mode.
We further note that, despite the reduced density of sample points along the H modes, we still compute $J_\mathrm{H}=0.064$ meV and $J_\mathrm{D}=4.9\times10^{-3}$ when the $T$ mode was excluded---consistent with our previously reported values for $J_\mathrm{H,D}$ \cite{Pritchard2025b}.
The ratio between our computed tunnel splittings, $J_\mathrm{D}/J_\mathrm{H}$, increases from 0.075 to 0.092 upon including the transition state mode $T$, indicating lattice distortions toward the transition state play an increasingly significant role for heavier tunneling species.

\autoref{fig:oh_vphi}a-i present contour plots of the potential energy surface and ground-state wavefunctions of O-trapped H along the $q_y$ and $T$ modes for two light-particle masses: $m=m_{\mu^+}$ and $m=m_\mathrm{H}$, with $q_x =q_z = 0$, and at three $Q$ coordinates: $Q_l$, $Q_s$, $Q_r$. 
As shown in \autoref{fig:oh_vphi}a-c, the primary role of the $Q$ mode is to symmetrize the minima of the double-well potential and reduce the barrier between degenerate hydrogen sites near transition-state configuration $T=T_{ts}$. 
For $m=m_\mathrm{H}$, the ground state wavefunction remains localized in the two adjacent, self-trapped  configurations at $Q=Q_l$ and $Q=Q_r$, with  suppressed population in the symmetric configuration $Q=Q_s$.
This behavior is consistent with earlier results and reflects the limited ability of H to access the barrier region  \cite{Pritchard2025b}.
Interestingly, the ground-state probability density for H is slightly biased toward the transition-state configuration, indicating that anharmonic lattice couplings are strong enough to shift the most probable hydrogen position away from the static minima.
\pgp{Nevertheless, as shown in \autoref{fig:oh_vphi}h, the most probable tunneling path does not approach the transition-state, indicating that tunneling does not occur along the minimum-energy path.}
In contrast to H, for $m=m_{\mu^+}$ the ground-state wavefunction is strongly biased toward the symmetric lattice configuration ($Q=Q_s$) and the transition-state distortion ($T=T_{ts}$); \pgp{the bias towards the transition-state persists} even in the self-trapped configurations ($Q=Q_{r,l}$).
This qualitative difference highlights the breakdown of adiabatic separation for H and D, but not for $\mu^+$.

Additional contours shown in \autoref{fig:oh_vphi}j-r illustrate the effect of the $T$ coordinate on the on $q_x$-$q_y$ plane ($q_z=0$ and \pgp{$Q=Q_s$}) for $T=0,$ $T_{ts}/2$,  and $T_{ts}$. 
Increasing $T$ lowers the magnitude of the local maximum in the potential energy at the octahedral position while simultaneously raising the potential energies of the local minima. 
For $m=m_{\mu^+}$, the tail of the wavefunction extends uniformly into the vicinity of the octahedral site for large $T$, suggesting a possible role for octahedral configurations.
As shown below, however, this contribution does not qualitatively alter the tunneling spectrum for the present geometry.

\subsection{Theories of light interstitials in bcc metals}

Theoretical descriptions of light interstitials in metals have long relied on the separation of time scales between electronic, lattice, and light-particle degrees of freedom.
In particular, models in which the light particle is treated adiabatically with respect to both electrons and host ions have been widely used to rationalize site preferences and tunneling behavior in bcc metals.
Here, we critically re-examine the validity of this adiabatic light-interstitial picture in bcc Nb by directly comparing it against the lattice-renormalized tunneling framework introduced above.

\subsubsection{Adiabatic light‑particle framework}

Sugimoto et al.\ introduced a theoretical framework in  which light interstitial particles, ranging from positive muons to tritium, are treated adiabatically with respect to both electrons \textit{and} host nuclei \cite{Sugimoto1980}.
The approach extends the Born-Oppenheimer approximation by introducing an intermediate separation of scales: electrons respond instantaneously to all ions, while the light particle evolves quantum mechanically in a potential defined by the average positions of the heavier lattice atoms.
In practice, the potential energy experienced by the light particle is sampled on a sufficiently large 3D grid surrounding its presumed equilibrium position, and the corresponding 3D Schr\"odinger equation is solved to obtain the light-particle wavefunction. 
The other ions comprising the host lattice are then relaxed under forces averaged over the light-particle probability density,
\begin{equation}
\label{eq:nonlocal_force}
\mathbf{F}_n=\int d\mathbf{r}\left|\psi(\mathbf{r})\right|^2\mathbf{f}_n(\mathbf{r})\,,
\end{equation}
where $\mathbf{F}_n$ is the wavefunction-weighted force acting on ion ${n}$, $\psi$ is the 3D wavefunction of the light particle, and $\mathbf{f}_n(\mathbf{r})$ is the force acting on ion ${n}$ when the light particle is fixed at position $\mathbf{r}$. \autoref{eq:nonlocal_force} may be derived from the Hellman-Feynman theorem. While not discussed in \pgp{Ref.} \cite{Sugimoto1980}, an analogous construction applies to stress by replacing $\mathbf{f}_n(\mathbf{r})$ with the local stress tensor $\sigma_{ij}(\mathbf{r})$.

Using an empirical two-body potential to evaluate $\psi$ and $\mathbf{f}_n$, Ref.~\cite{Sugimoto1980} concluded that H prefers tetrahedral interstitial sites in bcc Nb, consistent with later DFT studies \cite{Sundell2004,Pritchard2025b}, whereas, they conclude that positive muons preferentially favor octahedral sites. This contrasts with our finding that positive muons preferentially favor the symmetric configuration.
While this framework successfully captures certain qualitative trends, it rests on a strong assumption: that the light particle remains adiabatically decoupled from lattice dynamics even when tunneling between nearly degenerate configurations.

\subsubsection{Validity of the adiabatic approximation}

We now quantitatively assess the validity of the adiabatic light-particle approximation by explicitly testing its underlying separation of dynamical scales. 
Following Ref.~\cite{Sugimoto1980}, two nested Born-Oppenheimer (adiabatic) approximations are made, which we denote as the Light-Particle Approximation (LPA), yielding the hierarchy of Schr\"odinger equations shown below:

\begin{align}
\left[
-\sum_i \frac{\hbar^2}{2m_e}\nabla_i^2
+ V(\mathbf r_e;\mathbf r_p,\mathbf r_n)
\right]\psi_{e}^u
&=
E_e^u(\mathbf r_p,\mathbf r_n)\psi_e^u, \label{eq:bo-elec}
\\[4pt]
\left[
-\frac{\hbar^2}{2m_p}\nabla_p^2
+ E_e^u(\mathbf r_p;\mathbf r_n)
\right]\psi_p^v
&=
E_p^v(\mathbf r_n)\psi_p^v,\label{eq:bo-prot}
\\[4pt]
\left[
-\sum_i \frac{\hbar^2}{2m_i}\nabla_i^2
+ E_p^v(\mathbf r_n)
\right]\psi_n^w
&=
E_n^w\psi_n^w\,,\label{eq:bo-ion}
\end{align}
where semicolons denote variables treated parametrically and each wavefunction is understood to depend on the same variables appearing in the corresponding potential or effective energy term on the left hand side. 

\autoref{eq:bo-elec} is the electronic Schr\"odinger equation. We evaluate the ground state electronic energy, $E_e^0(\mathbf{r}_p, \mathbf{r}_n)$, at fixed ionic positions using DFT.
Here, $m_e$ is the mass of an electron, $\mathbf{r}_e$, $\mathbf{r}_p$, and $\mathbf{r}_n$ are the cartesian coordinates of the electrons, the light particle, and all other ions, respectively. 
\autoref{eq:bo-prot} is the Born-Oppenheimer Schr\"odinger equation of the light particle whose eigenstates, $\psi_p^v$, and eigenenergies, $E_p^v$,  depend parametrically on the position of all other ions. 
Finally, \autoref{eq:bo-ion} is the Born-Oppenheimer Schr\"odinger equation governing all remaining ions with eigenstates, $\psi_n^w$, and eigenenergies, $E_n^w$. The indices $v$ and $w$ index the eigenstates of the light-particle, and ionic Schr\"odinger equations.

\begin{figure}
\centering
   \includegraphics[width=0.38\textwidth]{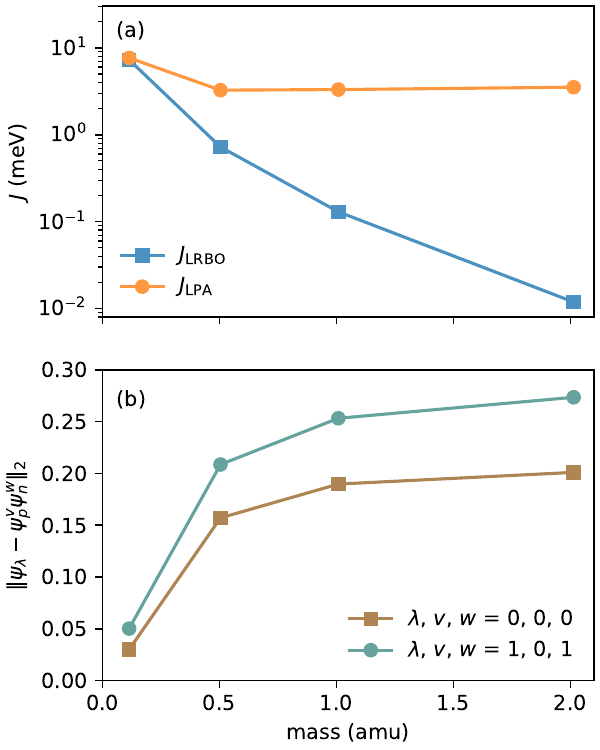}
   \vspace{-8pt}
   \caption{(a) Tunnel splittings ($J$) as a function of light-particle mass $m_p$, computed using the lattice-renormalized Born-Oppenheimer (LRBO) model and light particle approximation (LPA).
   (b) Deviation from wavefunction separability,
   $\epsilon_\psi = \|\psi_{\lambda} - \psi_p^v \psi_n^w\|_2$, of the LRBO ($\psi_\lambda$) and LPA  wavefunctions ($\psi_p\psi_n$), evaluated for ground and first-excited states.
   The LPA is satisfied only in the $\mu^+$ mass limit.}
   \label{fig:ad_fit}
\end{figure}

\begin{figure*}
\centering
   \includegraphics[width=0.74\textwidth]{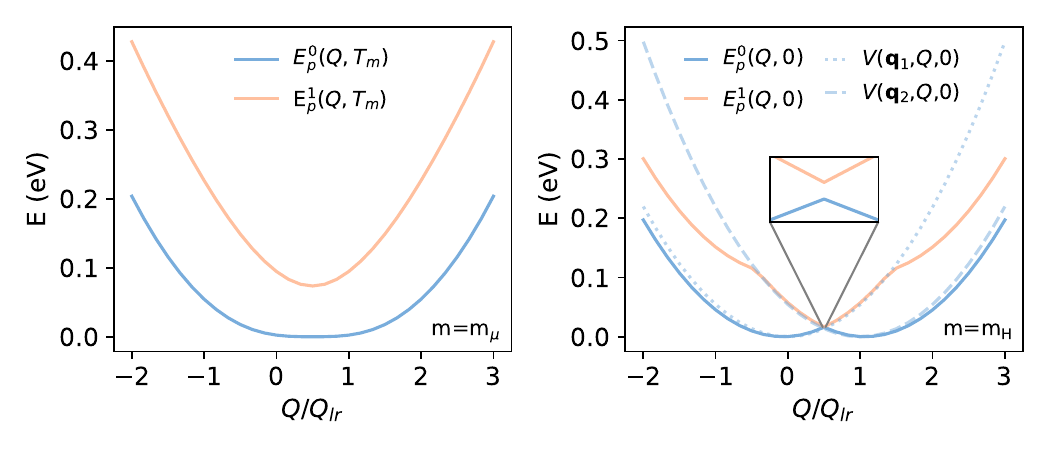}
   \vspace{-16pt}
   \caption{Adiabatic potential energy surfaces $E_p^v(Q,T)$ along the $Q$ coordinate at fixed $T$.
   (a) Ground state ($E_p^0$) and first-excited state ($E_p^1$) potential energy surfaces along the $Q$ mode for $m=m_\mu$ with $T$ set to its value at the local minimum $T_m$ shown in \autoref{fig:oh_vphi}. Well-separated energy levels characteristic of adiabatic behavior occur.
   (b) Corresponding surfaces for $m=m_\mathrm{H}$ with $T=0$, where only a small avoided crossing (inset) separates $E_p^0$ and $E_p^1$ near the symmetric configuration. 
   The LRBO potential energies $V(\mathbf{q}_1,Q,0)$ and $V(\mathbf{q}_2,Q,0)$ for hydrogen localized at the two degenerate sites along the $Q$ mode are also shown (see \autoref{fig:h_sites}), illustrating the breakdown of adiabatic level separation for hydrogen.}
   \label{fig:ad_energy}
\end{figure*}

The total wavefunction is then 
$$\psi_{uvw}(\mathbf{r}_e,\mathbf{r}_p,\mathbf{r}_n)=\psi_e^u(\mathbf{r}_e;\mathbf{r}_p,\mathbf{r}_n)\psi_p^v(\mathbf{r}_p;\mathbf{r}_n)\psi_n^w(\mathbf{r}_n)\,.$$
Assuming the conventional Born-Oppenheimer approximation for electrons is valid, the key question is whether this factorized form accurately reproduces the eigenstates of the full ionic Hamiltonian. Specifically, we test whether:
\begin{equation}
    \psi_{\lambda}(\mathbf{r}_p,\mathbf{r}_n)\approx\psi_p^v(\mathbf{r}_p;\mathbf{r}_n)\psi_n^w(\mathbf{r}_n)\,,
    \label{eq:psim}
\end{equation}
and if the resulting tunnel splitting $J=E_1-E_0$ is consistent with that obtained from the fully coupled light particle-lattice Hamiltonian. Here, $\psi_{\lambda}$ is the standard Born-Oppenheimer wavefunction for all ions in a given eigenstate, $\psi_p$ is the wave function of the light particle, $\psi_n$ is the wavefunction of all other ions and $\lambda$, $p$, $n$ index the eigenstates.

To perform this assessment, we recast Equations \ref{eq:bo-elec}--\ref{eq:bo-ion}  into our lattice-renormalized formalism where we exclude lattice modes assumed to behave harmonically.
We first transform the real space coordinates of the light-particle to phonon coordinates by standard methods \cite{ashcroft_solid_1976}.
Then, following the procedure for extracting out dominant lattice modes described in Ref.~\cite{Pritchard2025b} and above, we isolate a 2D subspace with only the $Q$ and $T$ modes, and obtain 
\begin{align}
\left[-\sum_i \frac{\hbar^2}{2m_e} \nabla_i^2 
+ V(\mathbf{r}_e;\mathbf{q},Q,T)\right]\psi_e^u 
&= E_e^u(\mathbf{q},Q,T)\psi_e^u, \label{eqBO1p} \\
\left[-\frac{\hbar^2}{2} {\nabla'}_p^2 
+ E_e^u(\mathbf{q};Q,T)\right]\psi_p^v 
&= E_p^v(Q,T)\psi_p^v,  \label{eqBO2p}\\
\left[-\sum_i \frac{\hbar^2}{2} {\nabla'}_i^{2} 
+ E_p^v(Q,T)\right]\psi_n^w
&= E_n^w \psi_n^w\,,
\label{eqBO3p}
\end{align}
so that the validity of the approximation becomes 
\begin{equation}
    \psi_{\lambda}(\mathbf{q},Q,T)\approx\psi_p^v(\mathbf{q};Q,T)\psi_n^w(Q,T)\,.
    \label{eq:psimp}
\end{equation}
To evaluate these approximations, we use the sampled potential energy surface, $E_e^0=V(\mathbf{q}, Q,T)$, and Equations~\ref{eqBO2p}--\ref{eqBO3p} to compute $E_n^w$, $\psi_p^v$, and $\psi_n^w$ for O-trapped H. 
We then obtain $\psi_{\lambda}$ as described in  Appendix~\ref{appendix:phisolve}.

\autoref{fig:ad_fit}a shows the tunnel splitting $J$ as a function of light-particle mass $m_p$, computed using both the lattice-renormalized Born-Oppenheimer ($J_\mathrm{LRBO}$) method and light-particle approximation ($J_\mathrm{LPA}$). 
While $J_\mathrm{LRBO}$ varies by nearly three orders of magnitude across the relevant mass range, 
$J_\mathrm{LPA}$ remains nearly constant and significantly overestimates $J_\mathrm{LRBO}$ except when $m\approx m_{\mu^+}$. 
\autoref{fig:ad_fit}b quantifies the  breakdown of the wavefunction separability through the $L_2$ norm given as $\epsilon_\psi=\|\psi_{\lambda}-\psi_p^v\psi_n^w\|_2$, evaluated for both the ground state ($\psi_\lambda=\psi_0$, $\psi_p^v\psi_n^w=\psi^0_p\psi^0_n$) and first-excited state ($\psi_\lambda=\psi_1$, $\psi_l\psi_n=\psi^0_p\psi^1_n$) wavefunctions.
\pgp{We note that the phase of $\psi_p^v(\mathbf{q};Q,T)$ may take an arbitrary value at each $Q$ and $T$ coordinate. 
We therefore impose a smooth gauge by requiring $\text{sign}(\psi_p^v(\mathbf{q}_1;Q,T)\psi_n^w(Q,T))=\text{sign}(\psi_\lambda(\mathbf{q}_1,Q,T))$.
}
We find that $\epsilon_\psi$ approaches zero only in the $m\approx m_{\mu^+}$ limit. 
These results demonstrate that the LPA is invalid for H and D interstitials in bcc Nb and becomes quantitatively reliable only for $\mu^+$.

\begin{figure*}
\centering
    \includegraphics[width=\textwidth]{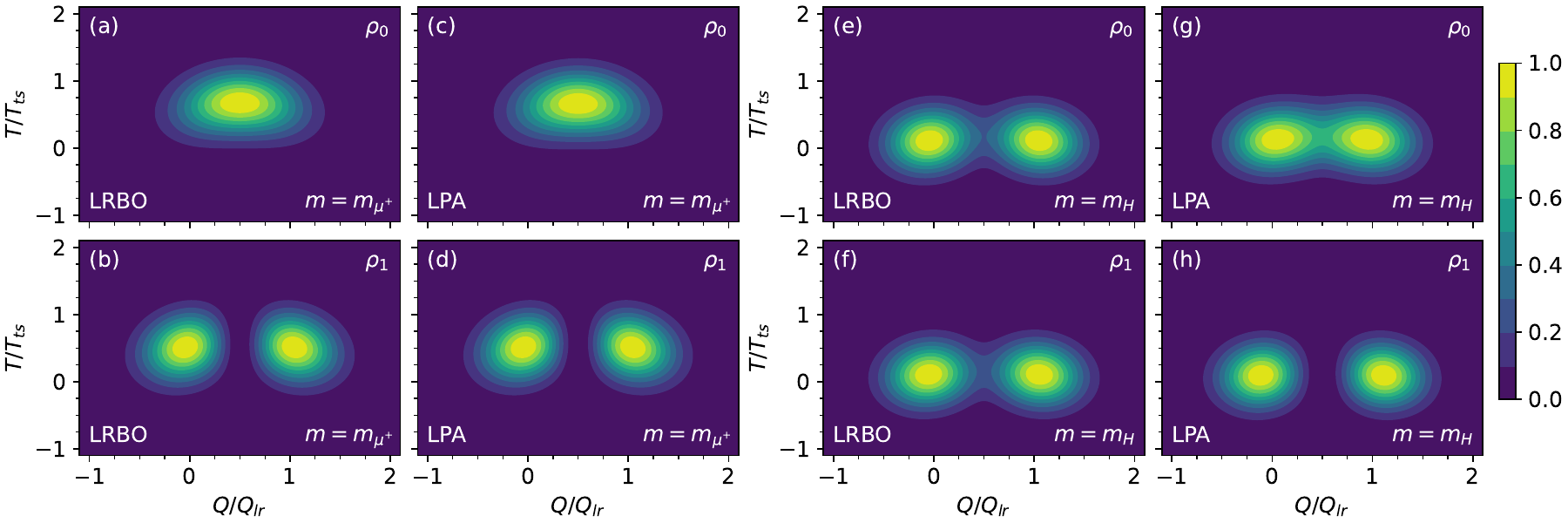}
   \caption{Lattice-mode probability densities $\rho_\lambda(Q,T)$ for the ground ($\rho_0$) and first-excited ($\rho_1$) tunneling states of O-trapped hydrogenic systems, computed using the LRBO (a, b, e, f) and LPA (c, d, g, h) formalisms. Panels (a, c, e, g) show  $\rho_0$, while (b, d, f, h) show $\rho_1$.
   For $m=\mu^+$, the two formalisms yield nearly identical lattice configurations, whereas for $m=m_\mathrm{H}$ the LPA predicts enhanced occupation of the symmetric configuration, leading to an overestimated tunnel splitting.}
   \label{fig:rho_lat_01}
\end{figure*}

\autoref{fig:ad_energy} further highlights these differences.
For $m=m_{\mu^+}$, the potential energy surfaces $E_p^0(Q,T)$ and $E_p^1(Q,T)$ are well separated, satisfying a fundamental requirement for adiabatic behavior.
In contrast, for $m=m_\mathrm{H}$, the two surfaces are separated only by a small avoided crossing near the symmetric configuration, reflecting strong nonadiabatic coupling. 
In this regime, the nominal ground state and first excited-state surfaces are approximately defined by $E_p^0=\min(V(\mathbf{q_1},Q,0),V(\mathbf{q_2},Q,0))$ and $E_p^1=\max(V(\mathbf{q_1},Q,0),V(\mathbf{q_2},Q,0))$ of the localized potential wells for a hydrogenic atom located at site 1 and site 2.
The first-excited state $E_p^1$ deviates from $V$ away from the symmetric configuration for large values of $|Q|$ as the energy asymmetry between wells exceeds the threshold for supporting a second local bound excited state in the lower energy well.

\subsection{Lattice configurations}

We now analyze how the mass of the light particle influences the most-probable lattice configurations associated with O-trapped hydrogenic tunneling systems. 
\autoref{fig:rho_lat_01} presents contours of the two-mode probability density 
\[
\rho_\lambda(Q,T) = \int d\mathbf{q}\, \left|\psi_\lambda(\mathbf{q},Q,T)\right|^2
\]
obtained by integrating out the light-particle modes $\mathbf{q}$ from the full 5D wavefunction $\psi_\lambda$. 
The quantity $\rho_\lambda(Q,T)$ provides direct information about which lattice distortions are preferentially occupied in a given tunneling eigenstate.
For $m=m_{\mu^+}$, the ground-state density $\rho_0$ is strongly localized near the symmetric configuration ($Q=Q_s$) and is significantly biased towards the transition-state distortion ($T= T_{ts}$), as shown in 
\autoref{fig:rho_lat_01}a.
This behavior \pgp{demonstrates} 
the strong tendency of the extremely light particle to delocalize between degenerate sites, thereby stabilizing lattice configurations near the barrier geometry.
In contrast, for $m=m_\mathrm{H}$, the ground-state lattice density exhibits equal weight near the left ($Q\approx Q_l$) and right ($Q\approx Q_r$) self-trapped configurations, with only a weak bias towards the transition state (\autoref{fig:rho_lat_01}e). 

A comparison between the LRBO and LPA formalisms further reveals a clear mass-dependence in the lattice-mode probability densities. 
For $m=m_{\mu^+}$, the ground ($\rho_0$) and first-excited-state ($\rho_1$) lattice densities are nearly indistinguishable (\autoref{fig:rho_lat_01}a-d), consistent with the small wavefunction mismatch $\epsilon_\psi$ discussed above.
This agreement indicates that the separation implicit in the LPA remains valid in the $\mu^+$ mass limit.
For $m=m_\mathrm{H}$, however, the two descriptions differ (\autoref{fig:rho_lat_01}e-h).
Within the LPA, $\rho_0$ is greatly enhanced at the symmetric configuration $Q=Q_s$ (\autoref{fig:rho_lat_01}g), whereas the corresponding LRBO result retains dominant weight near the self-trapped configurations (\autoref{fig:rho_lat_01}e).
This enhanced occupation of $Q_s$ within the LPA leads directly to the overestimation of the tunnel splitting shown in \autoref{fig:ad_fit}.
Differences between the two formalisms are further seen in the first-excited states.
The LPA predicts a clear nodal plane at $Q = Q_s$ in $\rho_1$, while the LRBO description exhibits only a modest suppression of lattice probability density at the symmetric configuration.
Thus, although both the LPA and LRBO approaches reproduce the expected two-state tunneling structure at low energies, they assign quantitatively different lattice participation to these states, resulting in substantially different tunneling energetics.

\begin{figure}
\centering
    \includegraphics[width=0.48\textwidth]{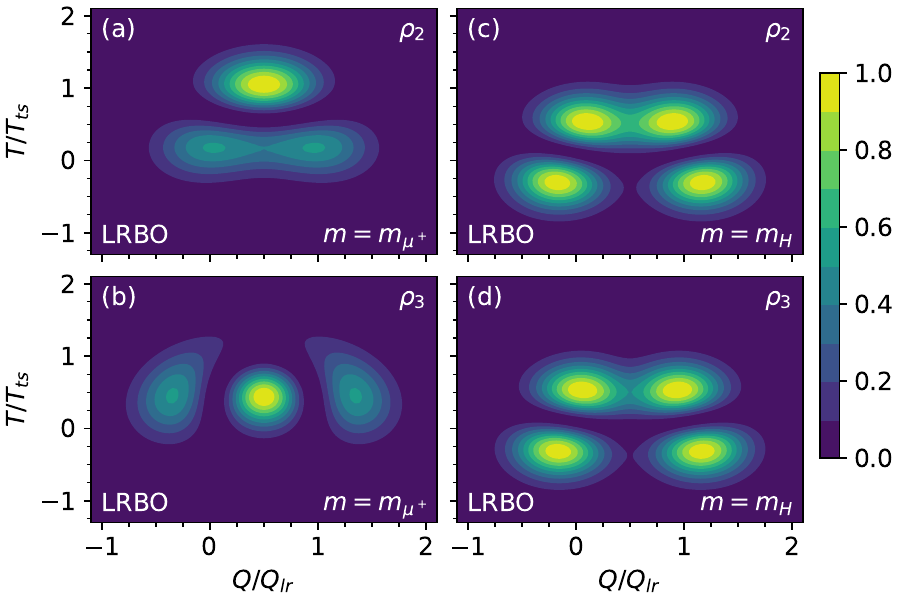}
   \caption{Lattice-mode probability densities for the second- ($\rho_2$) and third-excited ($\rho_3$) tunneling states computed using the LRBO. Panels (a, b) correspond to $m=m_{\mu^+}$, while (c, d) correspond to  $m=m_\mathrm{H}$.
   For $m=m_{\mu^+}$, the excited states are well described by harmonic excitations of the $T$ and $Q$ lattice modes. 
   For $m=m_\mathrm{H}$, the second-excited state exhibits mixed tunneling and lattice-vibrational character, reflecting nonadiabatic coupling.}
   \label{fig:rho_lat_23}
\end{figure}

The structure of the higher excited states also provide additional insight.
\autoref{fig:rho_lat_23} shows the second- ($\rho_2$) and third-excited state ($\rho_3$) lattice densities computed using the LRBO. 
For $m=m_{\mu^+}$, $\rho_2$ and $\rho_3$ may be clearly identified with the first-excited state of the $T$ mode and second-excited state of the $Q$ mode, respectively. Interestingly, $\rho_2$ exhibits partial delocalization across the local minima of the potential energy surfaces. 
For $m=m_\mathrm{H}$, the character of the second-excited state qualitatively changes. 
Here, $\rho_2$ exhibits hybrid character that combines tunneling biased towards the transition state $T\approx T_{ts}$ with an approximately harmonic first-excited state dominated by the $T$ coordinate.
These data support our conclusion that the excited-state structure cannot be described by a simple adiabatic hierarchy of lattice modes for hydrogenic interstitial species.

\subsection{Adiabaticity Criteria and Energetic Balance}
We now provide an energetic framework to rationalize the mass-dependent breakdown of the adiabatic LPA.
The underlying mechanism has two distinct but coupled contributions:
($i$) the energetic benefit associated with wavefunction delocalization, and 
($ii$) the anharmonic suppression of the tunnel barrier as the lattice is distorted toward the transition-state configuration (minimum barrier). 
For a two-level tunneling system associated with a local particle ($\Delta\gg J$), the ground-state energy is given by:
$$E_\mathrm{local}=-\frac{1}{2}\sqrt{\Delta^2+J_\mathrm{local}^2}+\frac{\Delta}{2}\,,$$
where $\Delta$ is the energy asymmetry between wells and $J_\mathrm{local}$ is the tunnel splitting associated with the local lattice configuration. 
In contrast, when the lattice adopts the symmetric configuration, the corresponding energy is 
$$E_\mathrm{sym}=-\frac{J_\mathrm{sym}}{2}+E_c\,,$$
where $J_\mathrm{sym}$ is the tunnel splitting in the symmetric configuration and $E_c$ is the elastic energy cost required to distort the lattice into that configuration. 
These expressions show that increasing the tunnel splitting $J$ lowers the symmetric-state energy more rapidly than the local-state energy.
Assuming that $J_\mathrm{local}\approx J_\mathrm{sym}$, the symmetric configuration becomes energetically favorable for large $J$ when $E_c < \Delta/2$. 
Moreover, if the light-particle wavefunction extends significantly across the transition state, local accommodation of the lattice occurs, further lowering the tunnel barrier and increasing $J_\mathrm{sym}$.
This additional reduction in kinetic energy is obtained at the expense of a higher elastic cost $E_c$.

For O-trapped H in bcc Nb with a concentration of $c=1/54$, the relevant parameters are $\Delta=54$\,meV and $E_c=12.5$\,meV \cite{Pritchard2025b}. 
When $m=m_{\mu^+}$, we compute $J_\mathrm{sym}=46$ meV. 
From the above energetic balance, we compute $J=36$\,meV as the minimum tunnel splitting required to stabilize the symmetric configuration, indicating that the symmetric lattice configuration is energetically preferred over the local configuration in this limit. 
More refined estimates may be obtained by directly computing the ground-state energy of the light-particle using a 3D Schr\"odinger equation at likely lattice configurations.  
We consider four lattice coordinates: 
$A=(0, 0)$,  
$B=(Q_s, 0)$, 
$C=(Qs, T_{ts}/2)$, and 
$D=(Q_s, T_{ts})$. 
The resulting energy differences are 
$\Delta E_{AB}=E_B-E_A=5.8$\,meV, 
$\Delta E_{AC}=-8.5$\,meV, and 
$\Delta E_{AD}=-7.5$\,meV. 
Interestingly, these data indicate that the symmetric configuration is \textit{not} energetically favored relative to the local configuration unless distortions along the transition-state mode are included explicitly.
This highlights the essential role of anharmonic lattice relaxation in stabilizing the symmetric configuration and enhancing tunneling in the extremely light-particle limit.

In summary, for O-trapped hydrogenic species in bcc Nb, the LPA introduced by Sugimoto et al.~\cite{Sugimoto1980} is only valid for particles as light as, or lighter than, $\mu^+$. 
In this limit, interwell tunneling is suppressed, and the lowest-energy transitions are between the lattice phonon modes whose reference state is defined on the adiabatic energy landscape of the $\mu^+$ particle. In contrast, hydrogen and deuterium tunnel nonadiabatically between adjacent minima, requiring a collective description that treats the light particle and lattice degrees of freedom on equal footing.

 \begin{table*}
 \label{tab:J_summmary}
     \centering
     \caption{Summary of reported computationally-derived tunnel splittings ($J_\mathrm{H}$, $J_\mathrm{D}$) for O-trapped H and D in bcc Nb. Sites are labeled following Ref.\  \cite{Pritchard2025b}. 
     Note that the H defect site used to compute  tunnel splittings in Ref.\ \cite{Abogoda2025a} is not a thermodynamically stable site and is equivalent to site $5a$ identified in Ref.\ \cite{Pritchard2025b}. The most stable O-trapped H site, site $7a$, was used in our works and in Ref.\ \cite{Abogoda2025b}.}
     \begin{ruledtabular}
     \begin{tabular}{llllllll}
           Site&Method&Hamiltonian&   Concentration&$J_\mathrm{H}$ (meV)& $J_\mathrm{D}$ (meV)  &$J_\mathrm{H}/J_\mathrm{D}$&Reference\\
 & & Dimension& & & & &\\
           \hline
  $5a$&NEB& 1D&  1/54&0.507& 0.0334  &15.2&\cite{Abogoda2025a}\\
  &VCNEB& 1D&  1/54&0.358& 0.0187  &19.1&\cite{Abogoda2025a}\\
  &Static lattice& 3D&  1/128&0.414& 0.024 &17.3&\cite{Abogoda2025b}\\
  \hline
            $7a$&Static lattice&3D&   1/128&0.275& 0.014 &19.6&\cite{Abogoda2025b}\\
  &Static lattice& 3D&  1/54&0.57& 0.040  &14.3&\cite{Pritchard2025b}\\
            &LRBO&4D&   1/54&0.064& 0.0048  &13.3&\cite{Pritchard2025b}\\
  &LRBO& 5D&  1/54&0.13&0.012 &10.8&This work\\
  \hline
 & Experiment& &  Several&0.19& 0.021 &9.0&\cite{Wipf1984p2}\\
     \end{tabular}
     \end{ruledtabular}
     
     \label{tab:placeholder}
 \end{table*}

\subsection{Comparison with Prior Tunnel-Splitting Calculations}
In addition to the present work and our prior work \cite{Pritchard2025b}, two other recent studies  have reported computationally-derived tunnel splittings for O-trapped H and D in bcc Nb using alternative methodologies \cite{Abogoda2025a, Abogoda2025b}. 
Abogoda et.~al computed the Minimum Energy Path (MEP) between adjacent tetrahedral sites using the Nudged Elastic Band (NEB) and Variable Cell Nudged Elastic Band (VCNEB) algorithms \cite{Abogoda2025a} with  energies and forces obtained from DFT simulations. 
The tunnel splittings for H and D were then computed using the potential energy along the reaction path of the MEP and a 1D Schr\"odinger equation with $m=m_\mathrm{H}$ or $m=m_\mathrm{D}$.  
Abogoda and Sauls also computed tunnel splittings by solving the three-dimensional Schr\"odinger equation for H in a static, symmetric bcc Nb lattice \cite{Abogoda2025b}. The potential-energy surface was obtained from a MACE machine-learned potential trained on DFT energies of O-H defect configurations.
\autoref{tab:J_summmary} compares tunnel splitting values reported in Refs.\ \cite{Abogoda2025a, Abogoda2025b, Pritchard2025b} and the present work. 
We find that $J_\mathrm{H}$ and $J_\mathrm{D}$ computed using a 1D Schr\"odinger equation along the MEP range  from 89\% to 167\% of the experimentally-reported values, depending on the mass and computational method used to compute the MEP.
The 1D Schr\"odinger equation defined along the MEP in Ref.\ \cite{Abogoda2025a}, however, neglects the effects of path curvature and correspondingly the mass of the tunneling particle is not well-defined.  As described above, the tunneling pathway computed using the LRBO formalism exhibits ``corner cutting" and hence does not reach the transition-state lattice configuration. This is consistent with the minimum action pathways computed for tunneling H (D) in the tropolone molecule using the instanton approach, which also exhibits ``corner cutting" \cite{Nandi2023}. %
Indeed, in tropolone, the instanton tunnel splitting along the MEP was computed to be approximately 500 times smaller than along the minimum action path. We therefore anticipate that the tunnel splitting computed using the MEP would be significantly suppressed if path effects were incorporated into the 1D model.

The most directly comparable tunnel splittings are those computed for site $7a$ in a static lattice using a 3D Schr\"odinger equation: $J_\mathrm{H}=0.275$ meV and $J_\mathrm{D}=0.014$ meV. As shown in \autoref{tab:J_summmary}, these values are lower than our 3D tunnel splitting values reported in Ref.\ \cite{Pritchard2025b} for the O-H defect, $J_\mathrm{H}=0.57$ meV and $J_\mathrm{D}=0.040$ meV. We note that the H concentration ($c=1/128$) was lower in Ref.\ \cite{Abogoda2025b} than in the present work ($c=1/54)$. We have previously shown that tunnel splittings computed using a static lattice overestimate the experimental splittings \cite{Pritchard2025b}, and that they are also approximately three to fourfold larger than those computed using the 5D LRBO Hamiltonian. We therefore expect that tunnel splittings computed using our 5D LRBO formalism combined with the MACE potential trained in Ref.\ \cite{Abogoda2025b} would be smaller than we report in the present work, and consequently in worse agreement with experiment.

\autoref{tab:J_summmary} also presents the ratio $J_\mathrm{H}/J_\mathrm{D}$ for each pair of tunnel splitting calculations. The experimental value of this ratio is $9.0$ while computational values range from 10.8 to 19.6. We note that the 5D LRBO result compares most favorably to the experimental tunnel splitting ratio.

\subsection{Implications for superconducting qubits}

Tunneling systems (TS) present in the bulk \cite{Zhang2024} and free surfaces of dielectric substrates \cite{Wang2015, Virginia2022}, as well as in the surface oxides of superconducting films \cite{Virginia2022, Bal2024}, are a dominant source of decoherence in superconducting qubits \cite{Mller2019}. 
Configurational tunneling systems (cTS), whose low-energy two-level or multi-level structure arises from degenerate or quasi-degenerate minima of the nuclear Hamiltonian, directly couple to the qubit through their electric dipole moment \cite{Martinis2005, Mller2019}. 
Within the conventional TLS framework, this coupling leads to resonant energy exchange when the transition energy matches the qubit frequency, as required by Fermi's golden rule.
Here, the minimum transition frequency is determined by the tunnel splitting $J$, and therefore precise knowledge of $J$ is essential to for identifying defects capable of resonant interaction.

For cTS embedded within superconductors, however, this decoherence channel is qualitatively altered.
Changes in the local dipole moment between adjacent configurations (minima) are screened by  superconducting electrons, suppressing the dominant coupling mechanism operative in dielectric environments. 
Nevertheless, as shown by He et al.~\cite{He2025}, a bath of tunneling TLS (such as those due to cTS) mediates additional interactions between quasiparticles in BCS superconductors.
These interactions depend on both the tunnel splitting $J$ and the distribution of well-to-well asymmetries, and give rise to competing effects:
an enhanced superconducting transition temperature (due to pair-enhancing interactions) and the formation of subgap quasiparticle (QP) states (due to pair-breaking interactions). 
In this regime, the conventional resonance condition is no longer the appropriate criterion for assessing whether a given cTS will induce additional decoherence;
rather, the relevant quantity is the induced distribution of subgap quasiparticle states, which depends on the full tunneling spectrum rather than on isolated resonant transitions.
Thus, defects with transition energies well above or below the qubit frequency \pgp{may} still contribute to decoherence. 
Our results for O-trapped H TLS demonstrate this as the tunnel splitting exceeds the typical transition frequencies of transmon qubits (deactivating the resonant mechanism), yet may still generates subgap QP states following \cite{He2025}. 
In contrast, O-trapped D, whose tunnel splitting  may be resonant with some qubit frequencies, is not expected to interact primarily through a resonant mechanism, but instead modifies the QP spectrum of subgap quasiparticle states similar to O-trapped H.
Thus, the classification of \pgp{cTS in superconductors} based on resonance conditions is insufficient; their impact should instead be understood through their contributions to the QP density of states.

\section{Conclusions}
We have developed a lattice-renormalized tunneling framework that captures the quantum dynamics of defect-trapped hydrogenic interstitials in bcc Nb within a reduced 5D Hamiltonian comprising three H (D) modes and two key lattice modes, $Q$ and $T$. 
The $Q$ mode connects adjacent minima of the nuclear Hamiltonian, while the $T$ mode drives the lattice toward the transition state from the symmetric configuration. 
Within this framework, We compute tunnel splittings of 0.13 meV (0.012 meV) for O-trapped H (D), in good agreement with the reported tunnel splittings \pgp{of 0.19 meV (0.021 meV)} inferred from specific heat and neutron-scattering experiments \cite{Wipf1984p2}. \pgp{The tunnel splitting ratio, $J_\mathrm{H}/J_\mathrm{D}=10.8$, computed using the 5D LRBO Hamiltonian also compares more favorably to experiment, $J_\mathrm{H}/J_\mathrm{D}=9.0,$ than our prior work \cite{Pritchard2025b} and other recent works \cite{Abogoda2025a, Abogoda2025b}.}
This level of accuracy is comparable to state-of-the-art instanton approaches \cite{Nandi2023}. 
Beyond quantitative agreement, this approach provides direct access to ground-state and excited-state wavefunctions of tunneling systems, confirming that the phenomenological two-level description \cite{Wipf1984, Wipf1984p2} of O-trapped H and D emerges from an underlying multilevel, lattice mediated tunneling process. 

By recasting the nested Born--Oppenheimer construction of Sugimoto et al.~\cite{Sugimoto1980} within this lattice-renormalized framework, we directly probe the validity of the adiabatic light-particle approximation.
We find that adiabatic separation of the light interstitial particle from the lattice only occurs for $\mu^+$ particles, whereas H and D exhibit intrinsically nonadiabatic behavior governed by strong anharmonic coupling to lattice modes.
This breakdown can be anticipated from simple energetic criteria based on the competition between tunneling-induced delocalization and the lattice distortion energy associated with transition-state configurations; it may be estimated by computing the energy of H sites in the self-trapped, coincidence, and transition state lattice configurations along with a single Schr\"odinger equation calculation for the light particle in a fixed lattice. 
These results establish that hydrogenic tunneling in bcc Nb is fundamentally a collective quantum process, requiring a treatment that places light-particle and lattice degrees of freedom on equal footing to accurately describe both tunneling energetics and their implications for superconducting materials.

\begin{acknowledgments}
This work was supported by the U.S. Department of Energy, Office of Science, National Quantum Information Science Research Centers, Superconducting Quantum Materials and Systems Center (SQMS), under Contract No.\  89243024CSC000002. Fermilab is operated by Fermi Forward Discovery Group, LLC under Contract No.\ 89243024CSC000002 with the U.S.\ Department of Energy, Office of Science, Office of High Energy Physics.
This research used resources of the National Energy Research Scientific Computing Center, a DOE Office of Science User Facility supported by the Office of Science of the U.S.\ Department of Energy under Contract No.\ DE-AC02-05CH11231 using NERSC award BES-ERCAP0036830.
This research was supported in part through the computational resources and staff contributions provided for the Quest high performance computing facility at Northwestern University which is jointly supported by the Office of the Provost, the Office for Research, and Northwestern University Information Technology. 
\end{acknowledgments}

\section*{Data Availability}
The data that support the findings of this article are openly available \cite{supp_repo}.
 
\section*{Appendix}
\appendix
\renewcommand{\thefigure}{A\arabic{figure}} %
\renewcommand{\thetable}{A\Roman{table}} 

\setcounter{figure}{0} 
\setcounter{table}{0} 

\section{Hamiltonian Solutions\label{sec:calcEq2}}

\subsection{Numerical evaluation\label{appendix:phisolve}}
We evaluate $V(\mathbf{q}, Q, T)$ on a sample grid spanning two degenerate H sites using density functional theory calculations.  
The orientation of the sample grid is identical to that used in Ref.~\cite{Pritchard2025b}; however, the sample grid dimensions along the H modes were augmented to fit the wavefunction of positive muons ($\mu^+$). Specifically, the hydrogen modes were defined on a grid $q_x, q_z\in \left[-q_{y_{12}}/2-q_{pad}, q_{y_{12}}/2+q_{pad}\right]$,  $q_y \in \left[-q_{y_{12}}/2-q_{pad},3q_{y_{12}}/2+q_{pad}\right]$; where, $q_{y_{12}}$ is the distance between the degenerate H sites in phonon coordinates and $q_{pad}$ was chosen such $m_\mathrm{H}^{-1/2}(q_{y_{12}}/2+q_{pad})\approx 0.875 $ \AA  \space while maintaining a uniform grid spacing. The $Q$ mode was sampled on the interval $Q\in\left[-2Q_{lr},3Q_{lr}\right]$ and the transition state mode was  sampled on the interval $T\in\left[-2T_{ts},3T_{ts}\right]$; $Q_{lr}$ and $T_ts$ denote the distances along $\mathbf{\hat Q}=\mathbf{Q}/|\mathbf{Q}|$ and  $\mathbf{\hat T}=\mathbf{T}/|\mathbf{T}|$; this space was sampled using $7\times11\times7\times11\times11$ evenly spaced grid points. \pgp{To simulate other hydrogenic species, $\mathbf{q}$ was rescaled by a factor of$\sqrt{m_p/m_\mathrm{H}}$.}

As before, we solve the Schr\"odinger on a subgrid whose spacing is an integer multiple of the sample grid's spacing. Cubic spline interpolation is used to map the DFT potential energy onto the subgrid. Eigenvalues and wavefunctions are obtained using Implicitly Restarted Lanczos Method implemented in \texttt{scipy}; all results are converged with respect to the subgrid resolution.

\subsection{DFT simulations}
All electronic structure simulations were performed with the Vienna Ab initio Simulation Package (VASP) \cite{Kresse1996, Kresse1999} and the Perdew-Burke-Ernzerhof (PBE) generalized gradient approximation (GGA) functional \cite{Perdew1996}. This functional has been previously shown to accurately capture the structure and elastic properties of bcc Nb \cite{Leibengood2025}. Projector augmented wave pseudopotentials \cite{Blochl1994} were used to treat core and valence electrons with the following electronic configurations: 4$s^2$4$p^6$5$s^1$4$d^4$ (Nb), 
2s$^2$2p$^4$ (O), and 1s$^1$ (H). The energy cutoff was 600 eV and the specified $k$ point grids were equal to $\lceil\frac{16}{n}\rceil\times\lceil\frac{16}{n}\rceil\times\lceil\frac{16}{n}\rceil$, where $n$ is the supercell dimension. Methfessel-Paxton smearing was used with $\sigma=0.2$ eV. 
\pgp{We have treated $\mu^+$ as a neutral defect and therefore no changes to DFT settings were required.}

\section{Coincidence structures\label{appendix:coincidence}}

In the adiabatic tunneling theory of Flynn and Stoneham \cite{Flynn1970}, the diffusion rate of a light interstitial atom is governed by an activation energy defined as the energy to form what is now called a ``coincidence'' configuration in which multiple minima of the nuclear Hamiltonian are degenerate. In their work, the coincidence structure was assumed to be equivalent to our symmetric configuration ($Q=Q_s$, $T=0$); however, as we have previously shown \cite{Pritchard2025b}, the energy of a coincidence configuration is lowered by relaxing orthogonal lattice modes.
The formalism of Ref.~\cite{Sugimoto1980} provides a convenient framework for constructing  such coincidence structures through wavefunction-weighted lattice forces.
In a coincidence geometry, the light-particle wavefunction is well approximated as a superposition of localized states centered at each potential minimum.
Within the harmonic approximation, the probability density and force fields may be expanded about each minimum $\mathbf{r}_i$ as
\begin{align*}
\left|\psi(\mathbf{r}-\mathbf{r}_i)\right|^2 &= \left|\psi(\mathbf{r}_i-\mathbf{r})\right|^2, \\
\mathbf{f}_n(\mathbf{r}-\mathbf{r}_i) &= \mathbf{f}_n(\mathbf{r}_i) + \Delta \mathbf{f}_n(\mathbf{r}-\mathbf{r}_i), \\
\mathbf{f}_n(\mathbf{r}_i-\mathbf{r}) &= \mathbf{f}_n(\mathbf{r}_i) - \Delta \mathbf{f}_n(\mathbf{r}-\mathbf{r}_i),
\end{align*}
where $\Delta \mathbf{f}_n$ denotes the linear change of the force along $\mathbf{r}-\mathbf{r}_i$. 
Under these approximations, the nonlocal force expression in \autoref{eq:nonlocal_force} reduces to the population-weighted average of forces evaluated at the most probable light-particle positions:
\begin{equation}
\label{eq:avg_force}
    \mathbf{F}_n=\left|\psi_1\right|^2 \mathbf{f}_n(\mathbf{r}_1)+\left|\psi_2\right|^2\mathbf{f}_n(\mathbf{r}_2)
\end{equation}
Where $|\psi_1|^2$ and $|\psi_2|^2$ are the integrated populations of wells 1 and 2. This expression is readily generalized to include an arbitrary number of minima. 
As the lattice is relaxed using Eq.~\eqref{eq:avg_force}, the positions $\mathbf{r}_i$ must be updated self-consistently to remain near the local extrema of the instantaneous potential energy surface.

In practice, structure relaxation with nonlocal light particle positions is performed by evaluating forces on the host (nonlight) atoms and the stress tensor using \autoref{eq:avg_force} (and its analog for stress).
We implement this scheme using a custom atomic simulation environment (ASE) compatible calculator and relax structures using the BFGS algorithm \cite{ASE}. 
In the special case of degenerate minima, \autoref{eq:avg_force} corresponds to equal population weights, 
$\left|\psi_1\right|^2 = \left|\psi_2\right|^2 = 0.5$,

For a two-level tunneling system described by a  double well potential with two minima, labeled 1 and 2, the ground-state energy may be written as:
\begin{equation}
E=-\frac{1}{2}\sqrt{\Delta^2+J^2}+\frac{V_1+V_2}{2}\,,
\end{equation}
where $V_1$ and $V_2$ are the local zero-point energies of each well, $\Delta=V_2-V_1$, and $J$ is twice the tunneling matrix element between the wells.
This expression may be generalized into an energy functional, $E(\mathbf{r}_n, \mathbf{r}_1, \mathbf{r}_2)$, which depends parametrically on the position of lattice ions, $\mathbf{r}_n$, excluding the light particle, and two light-particle positions, $\mathbf{r}_1$ and $\mathbf{r}_2$: 
%
\begin{equation}
\begin{aligned}
E(\mathbf{r}_n,\mathbf{r}_1,\mathbf{r}_2)
&=
-\frac{1}{2}
\sqrt{
    \Delta(\mathbf{r}_n,\mathbf{r}_1,\mathbf{r}_2)^2
    + J(\mathbf{r}_n)^2
}
\\
&\quad
+ \frac{
    V_1(\mathbf{r}_n,\mathbf{r}_1)
    + V_2(\mathbf{r}_n,\mathbf{r}_2)
}{2}\,,
\end{aligned}
\end{equation}
where $V_1$ and $V_2$ are defined in accordance with the potential energy of the Born-Oppenheimer nuclear Hamiltonian and are no longer restricted to the two minima. Specially, the image positions of the light particle are no longer assumed to be a local minimum. $\Delta=V_2(\mathbf{r}_n,\mathbf{r}_{2})-V_1(\mathbf{r}_n,\mathbf{r}_{1})$ is defined as the energy difference between each image position for fixed $\mathbf{r}_n$, and $J(\mathbf{r}_n)$ is  twice the tunneling matrix element.

The ground state populations of each well are given by:
\begin{align}\left|\psi_1\right|^2 &=\frac{1}{2}+\frac{1}{2}\frac{\Delta}{\sqrt{\Delta^2+J^2}},\\
\left|\psi_2\right|^2 &=\frac{1}{2}-\frac{1}{2}\frac{\Delta}{\sqrt{\Delta^2+J^2}}
\end{align}
To enforce the desired population constraints, we define the Lagrangian 
$$\mathscr{L}=E+\lambda\left(\left|\psi_2\right|^2-p_2\right)\,,$$
where $\lambda$ is a Lagrange multiplier and $p_2$ is the target population of well 2. 
A constrained minimum is achieved when the partial derivatives of $\mathscr{L}$ with respect to $\mathbf{r}_n$, $\mathbf{r}_1$, $\mathbf{r}_2$, and $\lambda$ are zero. By differentiation we obtain:
\begin{align*}
\mathbf{f}_{\mathscr{L}}&= -\left.\frac{\partial\mathscr{L}}{\partial \mathbf{r}_n}\right|_{\Delta=0}\\
&=- \frac{1}{2}\left(\frac{\partial V_1}{\partial \mathbf{r}_n}+\frac{\partial V_2}{\partial \mathbf{r}_n}\right)+\frac{1}{2}\frac{\partial J}{\partial\mathbf{r}_n}+\frac{\lambda}{2J}\frac{\partial\Delta}{\partial \mathbf{r}_n}\,=0,
\end{align*}
The first term in this expression is equivalent to the population-weighted average force (i.e.,  \autoref{eq:avg_force}) while the second term pulls the structure towards the configuration which maximizes the tunnel splitting. 
A similar force is present in \autoref{eq:nonlocal_force} when the wavefunction possesses a significant probability density near the barrier maximum, which activates the transition-state mode. 
When $J\approx0$, the potential energy function reduces to $E\approx\min(V_1,V_2)$. A local maximum is achieved when $V_1=V_2$ (i.e., the coincidence energy, $E_c$). This simultaneously satisfies the constraint $\Delta=0$, meaning that $\lambda=0$. For $J>0$, a local maximum of $E$ is not guaranteed to coincide with $\Delta=0$ in which case $\lambda\ne0$ may be required to extremize $\mathscr{L}$. In most practical situations, $J\ll E_c$, and therefore \autoref{eq:avg_force} may be used to compute coincidence structures.

\section{\label{sec:nonlocal} Adiabatic relaxation with $m=m_{\mu^+}$}

Our treatment above does not account for the effect of other lattice modes on the energy landscape of the light particle. Therefore, we applied \autoref{eq:nonlocal_force} to relax the Nb lattice for a $\mu^+$ trapped by an O interstitial ($c=1/16$) starting from a symmetric configuration to a force tolerance of $1\times10^{-2}$\,eV/\AA. We may then evaluate the suitability of our selected modes in capturing the adiabatically relaxed lattice. We define $\mathbf{R}^{ad}$ to be the center-of-mass coordinates of the structure relaxed using \autoref{eq:nonlocal_force}, excluding the $\mu^+$ particle, and $\Delta\mathbf{R}^{ad}=\mathbf{R}^{ad}-\mathbf{R}^l$ as the change in position from the self-trapped lattice configuration to the adiabatically relaxed lattice configuration. We convert $\Delta\mathbf{R}^{ad}$ to phonon coordinates by mass-normalizing, $\mathbf{Q}_{ad}=M^{1/2}\Delta\mathbf{R}_{ad}$. We then project the unit vector of this mode, $\mathbf{\hat Q}_{ad} $, onto $\mathbf{\hat Q}$  and $\mathbf{\hat T}$. We further define $c_Q= \mathbf{\hat Q}_{ad}\cdot\mathbf{\hat Q}$ and $c_T= \mathbf{\hat Q}_{ad}\cdot\mathbf{\hat T}$ and obtain $c_Q^2=0.32$ and $c_T^2=0.49$.  All  remaining (R) lattice modes contribute $c_\mathrm{R}^2=1-c_Q^2-c_T^2=0.19$ to the configuration of the adiabatically relaxed lattice. Therefore, the most probable lattice position is well-described by our two-mode model.

\bibliography{bib}
\end{document}